\documentclass[prl,superscriptaddress,twocolumn,showpacs]{revtex4}

\usepackage{enumerate}
\usepackage{amsfonts,amssymb,amsmath}
\usepackage[]{graphics,graphicx,epsfig}
\usepackage{amsthm}

\begin{document}

\title{Contextuality in bosonic bunching}

\author{Pawe\l\ Kurzy\'nski}
\email{cqtpkk@nus.edu.sg}
\affiliation{Centre for Quantum Technologies,
National University of Singapore, 3 Science Drive 2, 117543 Singapore,
Singapore}
\affiliation{Faculty of Physics, Adam Mickiewicz University,
Umultowska 85, 61-614 Pozna\'{n}, Poland}

\author{Akihito Soeda}
\email{cqtas@nus.edu.sg}
\affiliation{Centre for Quantum Technologies,
National University of Singapore, 3 Science Drive 2, 117543 Singapore,
Singapore}

\author{Jayne Thompson}
\email{cqttjed@nus.edu.sg}
\affiliation{Centre for Quantum Technologies,
National University of Singapore, 3 Science Drive 2, 117543 Singapore,
Singapore}

\author{Dagomir Kaszlikowski}
\email{phykd@nus.edu.sg}
\affiliation{Centre for Quantum Technologies,
National University of Singapore, 3 Science Drive 2, 117543 Singapore,
Singapore}
\affiliation{Department of Physics,
National University of Singapore, 3 Science Drive 2, 117543 Singapore,
Singapore}




\begin{abstract}
We show that under certain assumptions one can derive a variant of Specker's non-contextual inequality for a system of three indistinguishable bosonic particles. The inequality states that the sum of probabilities of three pairwise exclusive events is bounded by one. This inequality cannot be violated using standard quantum mechanical projectors. On the other hand, due to bosonic properties this bound is violated up to $3/2$. We also argue that the violation of this inequality can be considered as a test of bosonic nature.
\end{abstract}

\maketitle

{\it Introduction.}--- Contextuality is defined as a dependence of the measurement outcome on the choice of which other measurements are simultaneously performed. It was proven by Kochen and Specker (KS) \cite{KS} that any quantum system of dimension greater than two is contextual. 
Recently proposed experimental tests of contextuality are formulated in terms of Bell-like inequalities that are obeyed in every non-contextual theory and whose violation is considered as a manifestation of contextuality. The simplest non-contextual inequality is due to Klyachko-Can-Binicioglu-Shumovsky (KCBS) \cite{KCBS} who proved that for five cyclically exclusive events $\{A_1, \dots, A_5\}$ (by cyclically exclusive we mean that it is impossible for the pair of events $A_i$, and $A_{i+1}$, for any $i=1,\dots,5$ modulo 5, to happen at the same time) quantum mechanics does not allow joint probability distributions in accord with a non-contextual hidden variable model. KCBS derived an inequality for probabilities of these five events and showed that the sum of their probabilities cannot exceed $2$ for any non-contextual hidden variable theory 
\begin{equation}\label{KCBS}
\sum_{i=1}^5 p(A_i=1) \leq 2.
\end{equation}
It was also shown \cite{KCBS,Pent,C2} that in quantum mechanics the sum of probabilities for five cyclically orthogonal projective measurements can violate the bound of two, but can reach at most $\sqrt{5}$.

Interestingly, KCBS inequality belongs to a broader family of non-contextual inequalities based on $n$ cyclically exclusive events \cite{LSW,Ncycles}. The special case of $n=3$ is known as the Specker's inequality \cite{LSW,Specker} which states that the sum of probabilities of three pairwise exclusive events $\alpha$, $\beta$ and $\gamma$ is bounded by one 
\begin{equation}\label{Specker}
p(\alpha=1)+p(\beta=1)+p(\gamma=1)\leq 1,
\end{equation}
where $\alpha=1$ denotes occurrence of $\alpha$ while $\alpha=0$ means that $\alpha$ does not occur. This bound holds under the assumption that pairwise exclusiveness between all pairs of events implies mutual exclusivity of all events \cite{C2,IC2}. What is important, is that this assumption is obeyed by quantum mechanical projectors for which exclusivity is implemented via orthogonality relation, therefore it is commonly accepted that Specker's inequality can only be violated by theories that are more contextual than quantum mechanics.

Here, we present a physical system where, under certain assumptions, the sum of probabilities for three pairwise exclusive events exceeds the bound of 1 and reaches the value of $3/2$. Note, that $3/2$ is the maximal bound allowed by the no-disturbance principle \cite{G,CSW,Monogamy} which is a generalization of the no-signaling. This principle states that probabilities do not depend on the measurement context. More precisely, apart form exclusivity, the maximal value of $3/2$ results from the following assumptions
\begin{enumerate}
\item \label{compl} {\it Complementarity}: it is not possible to directly measure $p(\alpha=x,\beta=y,\gamma=z)$ ($x,y,z=0,1$); it is only possible to measure probabilities of pairs of events $p(\alpha=x,\beta=y)$, etc.
\item {\it Completeness}: $p(\alpha=0)+p(\alpha=1)=1$ (same for $\beta$ and $\gamma$)
\item {\it No-disturbance}: $\sum_y p(\alpha=x,\beta=y)=\sum_z p(\alpha=x,\gamma=z)=p(\alpha= x)$ (same for $p(\beta=y)$ and $p(\gamma=z)$) 
\end{enumerate}
Note, that in our case exclusivity demands that $p(\alpha=1,\beta=1)=p(\beta=1,\gamma=1)=p(\alpha=1,\gamma=1)=0$. The violation of the inequality (\ref{Specker}) can be maximized under the above assumptions by setting $p(\alpha=0,\beta=0)=p(\beta=0,\gamma=0)=p(\alpha=0,\gamma=0)=0$ and assigning $1/2$ to the remaining probabilities. As a result, $p(\alpha=1)=p(\beta=1)=p(\gamma=1)=1/2$ and the left hand side of the ineqality (\ref{Specker}) is $3/2$.

In quantum mechanics the event $\alpha$ is assosiated with the projector $\Pi_{\alpha}$ and the probability that $\alpha$ occurs is given by $\text{Tr}(\rho\Pi_{\alpha})$ for some state $\rho$. If $\alpha$ and $\beta$ are exclusive then the corresponding projectors are orthogonal, i.e., $\Pi_{\alpha} \cdot \Pi_{\beta}=0$. However if $\beta$ and $\gamma$ are also exclusive and if $\alpha$ and $\gamma$ are exclusive too, then the set of projectors $\Pi_{\alpha}$, $\Pi_{\beta}$, $\Pi_{\gamma}$ is mutually orthogonal and satisfies $\langle\Pi_{\alpha}\rangle+\langle\Pi_{\beta}\rangle+\langle\Pi_{\gamma}\rangle\leq 1$. This implies that inequality (\ref{Specker}) is satisfied for any quantum state. Moreover, what is important for the remaining discussion, mutual orthogonality of all three projectors implies that they are compatible, i.e., all of them can be measured in a single measurement which violates the above complementarity assumption \ref{compl} \cite{LSW}. 

\begin{figure}[t]
\begin{center}
\includegraphics[scale=0.4]{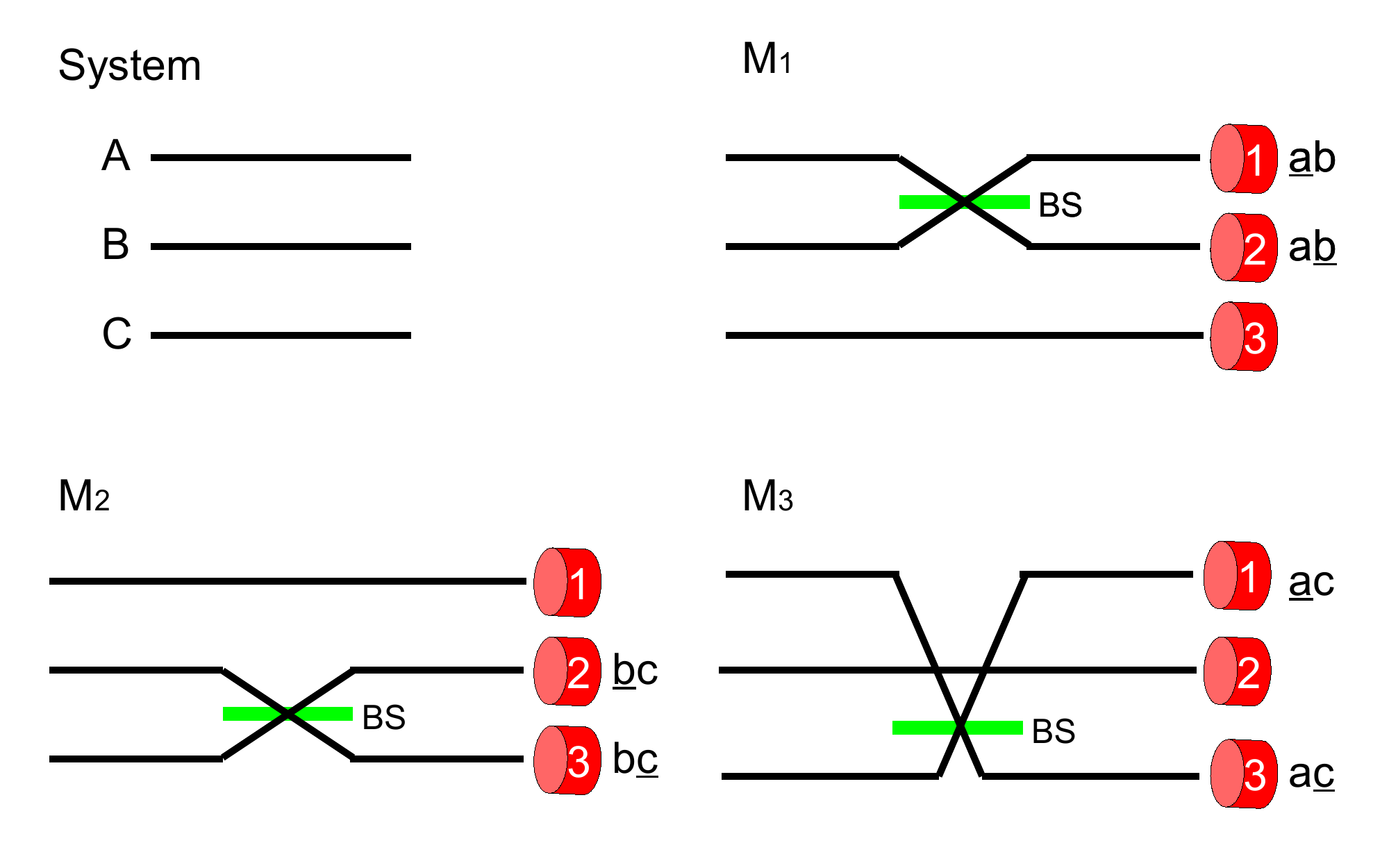}
\end{center}
\caption{\label{f1} Schematic picture representing the setup consisting of three photons in modes A, B and C (one photon per mode) and three different measurements $M_1$, $M_2$ and $M_3$. The measurements use a single 50/50 beam splitter (BS) to mix two modes and three detectors 1, 2 and 3.}
\end{figure}

We focus on a particular realisation of our bosonic system with three photons. Moreover, we utilize the bunching phenomenon which was demonstrated experimentally by Hong-Ou-Mandel (HOM) \cite{HOM}. Let us consider three photons in three optical fibers $A$, $B$, and $C$ (one photon per fiber). Next, let us consider three possible measurement scenarios $M_i$, $i=1,2,3$ that utilize a single beam splitter (BS) and three detectors that are pleaced at the end of each fiber. $M_1$ uses BS's inputs and outputs to mix modes corresponding to fibers A and B, $M_2$ mixes B and C and $M_3$ mixes A and C (see Fig. \ref{f1}). It is clear that every photon incident uppon the BS can be either reflected or transmitted. Furthermore, according to HOM \cite{HOM} photons will bunch, i.e., both photons from two inputs will exit together from a single output port (probability of $1/2$ for each output port).

{\it Assumptions and violation.}--- Let us introduce our assumptions upon which we will derive a variant of Specker's inequality (\ref{Specker}) for three bosonic particles   
\begin{enumerate}[i]
\item \label{md} {\it Mode distinguishability}: While bosons are still in fibers A, B and C, it is possible to refer to boson in fiber A as boson A, etc.
\item \label{nc} {\it Non-contextuality}: The scattering properties of each boson on the BS do not depend on which other fiber is connected to the other BS's input port and on the choice of the BS's input port
\item \label{r} {\it Realism}: It is possible to assign to each boson a binary variable $x$ ($x=a,b,c$ for bosons A, B and C, respectively) describing the scattering properties upon the BS, i.e., whether it is transmited (x) or reflected (\underline{x})
\end{enumerate}
The above assumptions are based on classical intuition. In addition, the assumption \ref{nc} is also supported by the fact that bosons do not interact while they are scattered by the BS. The BS's Hamiltonian includes only single-particle terms.

Next, consider the following three events: {\bf \underline{a}b} --- the photon A is reflected from BS AND B is transmited through BS; {\bf \underline{b}c} --- B is reflected AND C is transmited; {\bf a\underline{c}} --- C is reflected AND A is transmited. These events are composed of two elementary events refering to the behavior of a single photon. Moreover, they are pairwise exclusive. The events \underline{a}b and \underline{b}c are exclusive because b is exclusive to \underline{b}. Exclusivity of \underline{b}c and a\underline{c} and exclusivity of \underline{a}b and a\underline{c} folows from similar arguments.

Note, that events like \underline{a}bc (A is reflected {\it and} B is transmited {\it and} C is transmited), that include scattering of three photons, cannot be tested using our setup due to the fact that BS has only two input and two output ports. Interestingly, this fact is compliant with the complementarity assumption that was used to derive the bound of 3/2 for inequality (\ref{Specker}). Nevertheless, the assumptions \ref{md}, \ref{nc} and \ref{r} imply that one can construct a joint probability distributions over the space of eight events \{abc, \underline{a}bc, a\underline{b}c, ab\underline{c}, \underline{ab}c, \underline{a}b\underline{c}, a\underline{bc}, \underline{abc}\}. The existence of such a joint probability distribution implies that the following variant of inequality (\ref{Specker}) is obeyed:
\begin{equation}\label{bosonic}
p(\underline{a}b) + p(\underline{b}c) + p(a\underline{c}) \leq 1.
\end{equation}
This is because one can write $p(\underline{a}b)=p(\underline{a}bc)+p(\underline{a}b\underline{c})$, $p(\underline{b}c)=p(a\underline{b}c)+p(\underline{a}\underline{b}c)$, $p(a\underline{c})=p(ab\underline{c})+p(a\underline{b}\underline{c})$ and since the sum of all probabilities in the joint probability distribution is equal to one it is clear that 
\begin{equation}
p(\underline{a}b) + p(\underline{b}c) + p(a\underline{c}) = 1 - p(abc) - p(\underline{abc}) \leq 1. \nonumber
\end{equation}

However, inequality (\ref{bosonic}) can be violated by the setup presented in Fig. \ref{f1}. It is well known that scatering of two bosons on a 50/50 BS leads to the bunching phenomenon in which both bosons exit always together through one of the BS's output ports with probability 1/2 for each output \cite{HOM}. Due to this fact the probabilities $p(\underline{a}b)=p(\underline{b}c)=p(a\underline{c})=1/2$ which leads to the violation of inequality (\ref{bosonic}) since its left hand side is now $3/2$. This in turn implies that at least one of the assumptions \ref{md}, \ref{nc} and \ref{r} does not hold.

{\it Source of violation.}--- Inequality (\ref{bosonic}) would not be violated if the 
exclusivity of events \underline{a}b, \underline{b}c and a\underline{c} were implemented by pairwise orthogonality of three von Neumann projectors. However, as mentioned before, pairwise orthogonality of three projectors would imply their joint measureablity which does not occur in our case. We are going to show that in the case of indistinguishable particles there are events that are complementary but which at the same time can be considered as exclusive if one takes into account assumptions \ref{md}, \ref{nc} and \ref{r}. This complementarity is the source of the violation of inequality (\ref{bosonic}).

The measurements $M_i$ ($i=1,2,3$) use three detectors and a single BS. Despite the fact that the measurement is active in the sense that it contains a BS transformation, the measurement setup can be considered as a black box. Note, that in an idealized scenario the detector that is coupled to a fiber which is not connected to the BS will always click (for example, in $M_1$ detector 3 always clicks). Therefore, in each measurement only three detection events contain usefull information, i.e., both particles scattered by BS were detected by the upper detector (the particle from the upper mode was reflected and from the lower mode was transmited), both particles were detected by the lower detector (particle from the upper mode was transmited and from the lower mode was reflected), one particle was detected by the upper detector and one by the lower detector (however, we cannot say anything about which was reflected and which was transmited). As a result, the black box implementing the measurement $M_i$ contains three outputs corresponding to these three events.

Now, let us look at the problem from a different perspective using the Fock space approach. The initial state of the system can be expressed as $|1,1,1\rangle$, where the modes denote fibers A, B and C, respectively. On the other hand, the event $\underline{a}b$ corresponds to a projection onto a state $U_{BS}^{(1)\dagger}|2,0,1\rangle$, where $U_{BS}^{(1)\dagger}$ is the reversed BS transformation for the measurement $M_1$. Analogically, $\underline{b}c$ corresponds to projection onto $U_{BS}^{(2)\dagger}|1,2,0\rangle$ and $a\underline{c}$ to $U_{BS}^{(3)\dagger}|0,1,2\rangle$. We refer to these states as to $|\underline{a}b\rangle$, $|\underline{b}c\rangle$ and $|a\underline{c}\rangle$, respectively. The BS transformation transforms the creation operators of the upper ($u$) and lower ($l$) modes in the following way:
\begin{eqnarray}
a^{\dagger}_u & \rightarrow & \frac{a_u^{\dagger}+i a_l^{\dagger}}{\sqrt{2}}, \nonumber \\
a^{\dagger}_l & \rightarrow & \frac{i a_u^{\dagger}+ a_l^{\dagger}}{\sqrt{2}}. \nonumber
\end{eqnarray}        
It is therefore straightforward to show that
\begin{eqnarray}
|\underline{a}b\rangle=\frac{1}{2}(-i\sqrt{2}|1,1,1\rangle + |2,0,1\rangle - |0,2,1\rangle), \nonumber \\
|\underline{b}c\rangle=\frac{1}{2}(-i\sqrt{2}|1,1,1\rangle + |1,2,0\rangle - |1,0,2\rangle), \nonumber \\
|a\underline{c}\rangle=\frac{1}{2}(-i\sqrt{2}|1,1,1\rangle + |0,1,2\rangle - |2,1,0\rangle). \nonumber
\end{eqnarray} 
It is clear, that in the Fock space representation the three events are complementary, since $|\langle \underline{a}b| \underline{b}c\rangle|^2=|\langle \underline{b}c| a\underline{c}\rangle|^2=|\langle \underline{a}b| a\underline{c}\rangle|^2=1/4$.

The exclusivity of the three events that enter inequality (\ref{bosonic}) is not physicaly testable. It rather stems from the assumptions \ref{md}, \ref{nc} and \ref{r}, and from the fact that each event is a composition of two single-photon events. 
This resembles the exclusivity of composite events discussed in \cite{C2}, where exclusive events like (p AND q) and (\underline{p} AND r) were defined for two independent experiments (\underline{p} is exclusive to p). Note, that events p and \underline{p} were exclusive events in one laboratory, whereas q and r were some events in the other laboratory. However, the events in each laboratory were represented by projectors $\Pi_p$, $\Pi_{\underline{p}}$, $\Pi_q$, $\Pi_r$ and composite events were defined as a tesnor product of projectors corresponding to different laboratories. What is important, is that the tensor product structure takes care of compatibility, because although $\Pi_q$ and $\Pi_r$ may not be orthogonal, the projectors $\Pi_p\otimes \Pi_q$ and $\Pi_{\underline{p}}\otimes \Pi_r$ are orthogonal due to orthogonality of $\Pi_p$ and $\Pi_{\underline{p}}$. The reason why our case is different is that a tensor product structure does not naturaly occur for indistinguishable particles which is the root of complementarity of the events that are assumed to be exclusive.

In the beginning we showed that the violation of the Specker's inequality (\ref{Specker}) up to $3/2$ is possible under the assumptions of complementarity, completeness and no-disturbance. The above arguments show that our model obeys the complementarity assumption and it is clear that all probabilities in the experiment fulfill the completeness assumption. Moreover, it is easy to show that the no-disturbance assumption is also valid. Note that the probability that {\it a particular} photon is reflected (transmited) does not depend on which other photon enters the other BS's input port. For example, the probability that photon A is reflected is the same independent of whether it is scattered together with photon B or C 
\begin{equation}
p(\underline{a})=p(\underline{a}b)+p(\underline{ab})=p(\underline{a}c)+p(\underline{ac})=1/2. \nonumber
\end{equation}
Note, that in our case the no-disturbance is intertwined with the indistinguishability. Since photons are indistinguishable, the probabilities of reflection or transmission cannot depend on the choice of the photon in the other port.

{\it KCBS-like scenario.}--- An analogical approach can be used to formulate an alternative version of the KCBS-like scenario with five cyclically exclusive events. This time consider five photons in five optical fibers A, B, C, D and E. In Fock space representation the state of the system is of the form $|1,1,1,1,1\rangle$. The five measurement scenarios utilize five detectors coupled to each fiber and a single BS that mixes modes A and B ($M_1$), B and C ($M_2$), C and D ($M_3$), D and E ($M_4$), or A and E ($M_5$). The five cyclically exclusive events are: {\bf \underline{a}b}, {\bf \underline{b}c}, {\bf \underline{c}d}, {\bf \underline{d}e}, and {\bf a\underline{e}}. Again, due to the bunching phenomenon the probability of each event is $1/2$ and hence the KCBS inequality is violated up to 5/2. It can be also shown that the events which are considered to be exclusive are also complementary.

{\it Discussion.}--- Because the above results seem to be contradictory to the recent proof by Cabello \cite{C2} that exclusivity forbids the violation of the KCBS inequality to be greater than $\sqrt{5}$, it was argued that our bosonic schemes do not test contextuality \cite{Cabello2}. However, we argue that there is no contradiction at all, because the contextuality discussed in this work differs from the one that is usually tested by noncontextual inequalities and that our scheme tests contextuality of a different type than the one defined by KS. 

In Ref. \cite{Cabello2} it was argued that an experiment that tests some noncontextual inequality should have the following properties: (a) all measurements are performed on a system in the same state, (b) experiments should involve only compatible (repeatable) tests, (c) each test has to appear in more than one set of different compatible tests. Our scheme satisfies the first condition, since the state of the system on which a measurement is performed is always the same. In the case of inequality (\ref{bosonic}) it is $|1,1,1\rangle$ and in the case of the KCBS-like scenario it is $|1,1,1,1,1\rangle$. However, the last two conditions are not fulfilled. 

Due to the indistinguishable nature of particles compatibility and repeatibility do not occur in our proposal. After the scattering event the two photons cannot be distinguished. Note, that this problem also occurs in other types of contextuality. For example, contextuality using generalized measurements (POVMs) \cite{CPOVM} also involves tests that are not repeatable \cite{CPOVM2}. Moreover, the notion of contextuality, presented in this work, refers to the fact that one can choose whether to scatter photon A with B ($M_1$) or with C ($M_2$) and to the fact that it is not possible to assign properties to individual bosons independently of this choice. We would like to highlight that the above notion of contextuality does not mean the multiplicity of measurement contexts for the two-photon events that we are testing.

{\it Outlook.}--- The HOM experiment is often considered as a test of bosonic nature, however note that the bunching phenomenon between two photons on a single BS can be explained using the outcome assignment model presented in this work. If one assigned values (transmited/reflected) to two distinguishable particles A and B it would be possible to simulate bunching statistics. One simply assigns $p(a\underline{b})=p(\underline{a}b)=1/2$. On the other hand, the addition of the third particle and the ability to make a choice which two particles to send on a BS results in the inequality (\ref{bosonic}), that itself can be considered as a more rigorous test of bosonic nature. We conjecture that in a similar way it is possible to extend our result to create more rigorous tests of fermionic nature. Note, that other new tests of indistinguishability have been recently proposed in \cite{Bose}.

Bosonic effects attract much attention due to the new idea of boson sampling \cite{BS} in which particle statistics is applied to solove problems that cannot be efficiently solved using classical resources. It is therefore natural to ask whether the power of boson sampling is related to the contextuality discussed in this work. If this is the case, boson sampling would be a powerful application of this new type of contextuality and one may hope to extend it further to contextuality of the KS type. Moreover, we are currently able to amplify randomness using two local boxes \cite{Rand}. However, it would be more practical if we had only one box for this purpose. It is argued that boson sampling can be simulated classically for all practical purposes \cite{Eisert}. If we could use our test to guarantee lack of classical simulation then we will have a quantum box doing boson sampling and therefore producing quantum random numbers. 

{\it Acknowledgements.}--- P. K. acknowledges helpful discussions with Andrzej Grudka and Antoni W\'ojcik. This work is funded by the Singapore Ministry of Education (partly through the Tier 3 Grant "Random numbers from quantum processes") and by the Singapore National Research Foundation. P. K. was also supported by the Foundation for Polish Science.


\begin{thebibliography}{99}

\bibitem{KS}
S. Kochen and E.P. Specker,
J. Math. Mech., 17, {\bf 59} (1967).

\bibitem{KCBS}
A. A. Klyachko, M. A. Can, S. Binicioglu, and A. S. Shumovsky, 
Phys. Rev. Lett. {\bf 101}, 020403 (2008).

\bibitem{C2}
A. Cabello,
Phys. Rev. Lett. {\bf 110}, 060402 (2013).

\bibitem{Pent}
P. Badziag, I. Bengtsson, A. Cabello, H. Granstrom, J.-A. Larsson,
Found. Phys. {\bf 41}, 41 (2011).

\bibitem{LSW}
Y.-C. Liang, R. W. Spekkens, and H. M. Wiseman,
Phys. Rep. {\bf 506}, 1 (2011).

\bibitem{Ncycles}
M. Araujo et. al., arXiv:1206.3212 (2012).

\bibitem{Specker}
E.P. Specker,
Dialectica {\bf 14}, 239 (1960).

\bibitem{IC2}
J. Henson,
arXiv:1210.5978 (2012).

\bibitem{G}
A. M. Gleason, 
J. Math. Mech. {\bf 6}, 885 (1957).

\bibitem{CSW}
A. Cabello, S. Severini, and A. Winter,
arXiv:1010.2163 (2010).

\bibitem{Monogamy}
R. Ramanathan, A. Soeda, P. Kurzynski, and D. Kaszlikowski,
Phys. Rev. Lett. {\bf 109}, 050404 (2012).

\bibitem{HOM}
C. K. Hong, Z. Y. Ou, and L. Mandel,
Phys. Rev. Lett. {\bf 59}, 2044 (1987).

\bibitem{Cabello2}
A critique of the early version of this manuscript (arXiv:1211.6907) was presented in
A. Cabello,
arXiv:1212.5502 (2012).

\bibitem{CPOVM}
A. Cabello,
Phys. Rev. Lett. {\bf 90}, 190401 (2003).

\bibitem{CPOVM2}
A. Grudka and P. Kurzynski,
Phys. Rev. Lett. {\bf 100}, 160401 (2008).

\bibitem{Bose}
S. Bose and D. Home,
Phys. Rev. Lett. {\bf 110}, 140404 (2013).

\bibitem{BS}
S. Aaronson and A. Arkhipov,
Proceedings of ACM Sympoium on the Theory of Computing, STOC (2011).

\bibitem{Rand}
R. Colbeck and R. Renner,
Nat. Phys. {\bf 8}, 450 (2012). 

\bibitem{Eisert}
J. Eisert, private communication.



\end{thebibliography}
\end{document}